\documentclass{IEEE_lsens}
\usepackage{multirow}
\usepackage[table,xcdraw]{xcolor}

\usepackage[noadjust]{cite}

\ifCLASSINFOpdf
  \usepackage[pdftex]{graphicx}
\else
\fi
\usepackage[T1]{fontenc} 
\usepackage{amsmath}
\usepackage[cmintegrals]{newtxmath}
\usepackage{bm}

\usepackage{array}

\usepackage[caption=false,font=footnotesize,labelfont=sf,textfont=sf]{subfig}
\usepackage{url}
\ifCLASSINFOpdf
\else
\fi
\providecommand{\hypersetup}[1]{\relax}

\hypersetup{pdftitle={PreMovNet: Pre-Movement EEG-based Hand Kinematics Estimation for Grasp and Lift Task},
pdfsubject={Typesetting},
pdfauthor={Anant Jain},
pdfkeywords={Class, IEEE, IEEE\_lsens, IEEE Sensors Letters, LaTeX, Typesetting, TeX}}
\usepackage{color}

\begin{document}


\IEEELSENSarticlesubject{IEEE Sensor Letters}

%
\title{PreMovNet: Pre-Movement EEG-based Hand Kinematics Estimation for Grasp and Lift Task}

%
\author{\IEEEauthorblockN{Anant~Jain\IEEEauthorrefmark{1},and~Lalan~Kumar\IEEEauthorrefmark{1,2}
}
\IEEEauthorblockA{\IEEEauthorrefmark{1}Department of Electrical Engineering,
Indian Institute of Technology Delhi, New Delhi 110016 India\\
\IEEEauthorrefmark{2}Bharti School Of Telecommunication Technology And Management,
Indian Institute of Technology Delhi, New Delhi 110016 India
}%
\thanks{This work was supported in part by DRDO - JATC project with project number RP04191G.}
}
%
%
%


\IEEEtitleabstractindextext{%
\begin{abstract}
Kinematics decoding from brain activity helps in developing rehabilitation or power-augmenting brain-computer interface devices. Low-frequency signals recorded from non-invasive electroencephalography (EEG) are associated with the neural motor correlation utilised for motor trajectory decoding (MTD). In this communication, the ability to decode motor kinematics trajectory from pre-movement delta-band (0.5-3 Hz) EEG is investigated for the healthy participants. In particular, two deep learning-based neural decoders called PreMovNet-I and PreMovNet-II, are proposed that make use of motor-related neural information existing in the pre-movement EEG data. EEG data segments with various time lags of 150 ms, 200 ms, 250 ms, 300 ms, and 350 ms before the movement onset are utilised for improving the decoding performance of the neural decoders. The MTD is presented for grasp-and-lift task (WAY-EEG-GAL dataset) using EEG with the various lags taken as input to the neural decoders. The performance of the proposed decoders are compared with the state-of-the-art multi-variable linear regression (mLR) model. Pearson correlation coefficient and hand trajectory are utilized as performance metric. The results demonstrate the viability of decoding 3D hand kinematics using pre-movement EEG data, enabling better control of BCI-based external devices such as exoskeleton/exosuit.
\end{abstract}

\begin{IEEEkeywords}
brain computer interface (BCI), electroencephalography (EEG), multi-variable linear regression (mLR), deep learning, pre-movement.
\end{IEEEkeywords}}


\maketitle

\section{Introduction}
\IEEEPARstart  
Brain-computer interface (BCI) incorporates brain activity to control external devices. In particular, Electroencephalography (EEG) signals are commonly utilised due to its non-invasive nature, mobility, and low cost. EEG signal has been widely utilized for motor classification \cite{zhang2019classification,chaisaen2020decoding}, commanding robotic devices \cite{mishchenko2018developing} and instinctual control of prosthetic devices based on continuous classification \cite{gao2019eeg}. However, it is to note that continuous kinematic parameter prediction based motor trajectory decoding (MTD) would provide better control of the external devices such as neural prosthetic, exosuit or exoskeleton devices. Recent literature supports EEG-based MTD for upper-limb in offline\cite{bradberry2010reconstructing, Sosnik2020, pancholi2022source,jeong2019trajectory}, and online mode\cite{jeong2020brain}. The technique utilized for this purpose is multi-variable linear regression (mLR) \cite{bradberry2010reconstructing, Sosnik2020}. However, mLR technique is based on linear relationship between input-output variables and is sensitive to outliers. This shortcoming of mLR technique can be overcome by deep learning based decoding models. Deep learning models utilize non-linear features and are robust to the outliers. Deep learning based frameworks have been used for MTD \cite{jeong2019trajectory, jeong2020brain, pancholi2022source} using EEG signals. Convolutional neural network (CNN)-Bidirectional long short-term memory (BiLSTM) based arm trajectory decoding for robotic arm control, was established using EEG in \cite{jeong2020brain} with an average correlation reported as 0.47. Visual stimuli was considered as reference point in \cite{pancholi2022source} for CNN-LSTM based trajectory estimation. However, visual stimuli may not be always present in real world applications. Hence, movement onset as reference point is explored for MTD in this work. In particular, delta frequency band is utilized as input features to neural decoder for efficient decoding \cite{ofner2012decoding, ubeda2015assessing}.

In this study, two neural decoding models, PreMovNet-I (MLP based) and PreMovNet-II (CNN-LSTM based), are proposed for robust MTD. It makes use of motor-related neural information existing in the pre-movement EEG data. It may be noted that the motor movement is encoded in the EEG signal around 300 ms prior to actual movement \cite{pancholi2022source}. Hence, EEG with an appropriate lag from the movement onset is taken as input to neural decoders. The inclusion of this neural information helps improve the performance of MTD. In particular, 3D hand trajectory decoding is performed for the grasp-and-lift task \cite{luciw2014multi}. Twelve right-handed participants' data sets were included. For the inclusion of pre-movement neural information, EEG lags of 150 ms, 200 ms, 250 ms, 300 ms, and 350 ms from movement onset were taken for evaluating the performance of the proposed neural decoder for 3D hand movement. Thus, the input to the neural decoder is EEG with time lags. The channels were selected around the motor-cortex region, which is mapped to the hand movement. EEG signal recordings are initially preprocessed for noise removal and filtered in the delta frequency band (0.5–3 Hz). The 3D hand position kinematics data is also preprocessed prior to neural decoder training and testing. Both EEG and kinematics data have been down-sampled to 100 Hz to reduce the computation complexity of the neural decoder. The proposed deep learning frameworks, PreMovNet-I and PreMovNet-II, are compared with state-of-the-art mLR for the reach-and-grasp task.
\begin{figure*}[!t]
	\centering
	\captionsetup{justification=centering}
	\includegraphics[width=0.6\textwidth]{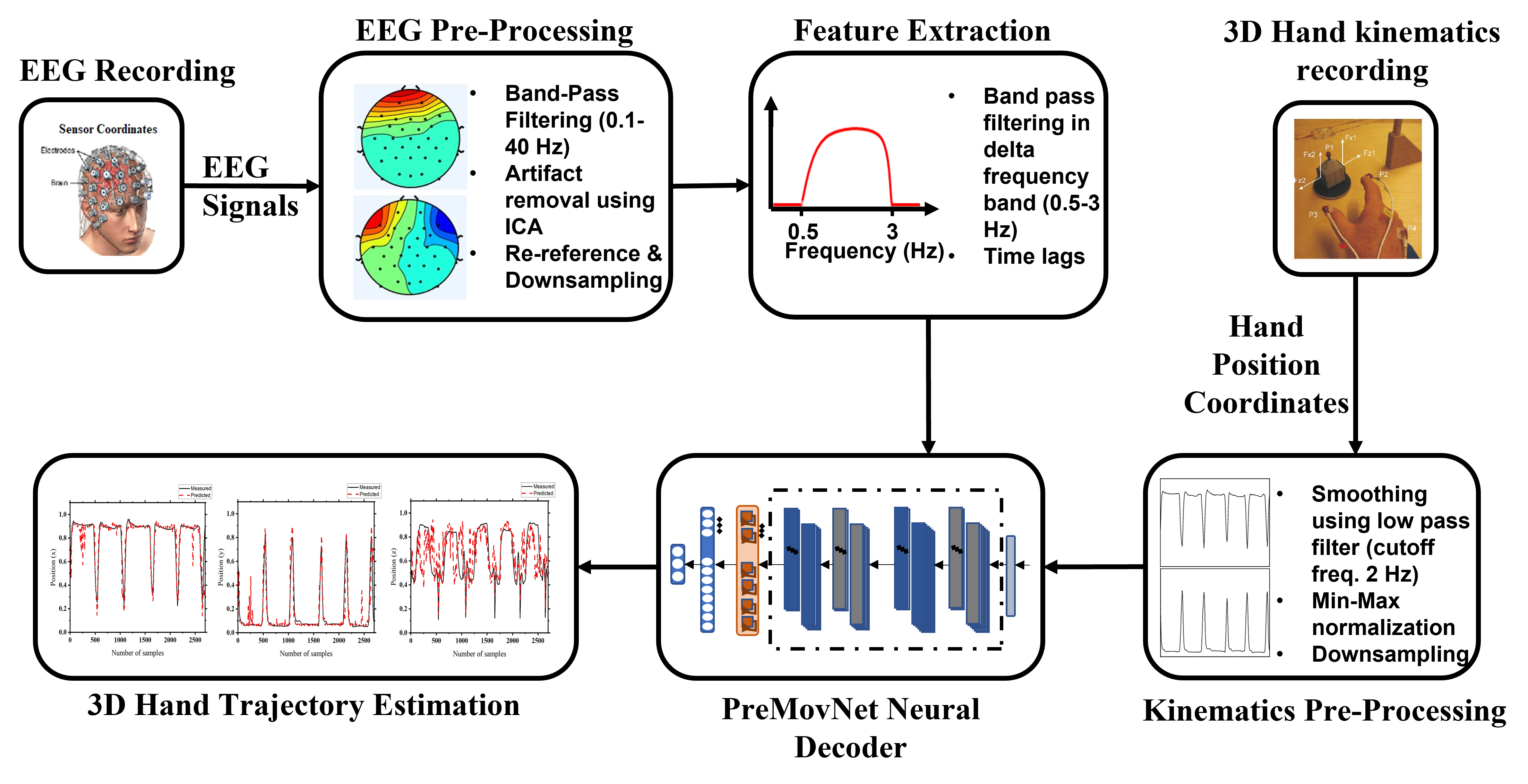}
	\caption{Flowchart of pre-movement EEG based hand kinematics estimation framework.}
	\label{01_flowchart}
\end{figure*}

\section{Methodology}
In this Section, an explicit description of the utilised dataset, data preprocessing steps, and EEG-based kinematics estimation framework is presented. Fig. \ref{01_flowchart} illustrates the process flow of the proposed hand kinematics estimation framework.
\subsection{Experimental Setup}
The WAY-EEG-GAL database \cite{luciw2014multi} is utilized for the performance evaluation of the MTD framework with the kinematics data of twelve subjects. A 32-channel ActiCap (Brainproducts) was utilized to record the EEG data and wrist position (XYZ Cartesian coordinates) of the subjects was recorded using a 3D position sensor with a sampling rate as 500 Hz. The subjects performed a series of grasp and lift task with various loads (165, 330, or 660 g) and surface frictions (silk, suede, or sandpaper).

Each trial began when the LED was turned on. The participant reached for the object, grasped it with forefinger and thumb, and then lifted it in the space stably for two seconds. Upon turning off the LED, the participant had to lower the hand, put the object in the original position and retracting his or her arm to its initial resting position.

\vspace{-0.15cm}

\subsection{Data Pre-Processing \& Feature Extraction}
The recorded EEG data was first band-pass filtered using a zero-phase FIR filter in the frequency band of 0.1–40 Hz in order to discard baseline drifts. The data was then re-referenced to the average of all scalp electrodes.Then, the eye movement artefacts were removed using the independent component analysis (ICA) technique. The processed EEG data was further down-sampled to 100 Hz to reduce the computational cost. All the EEG data preprocessing was performed using the EEGLAB \cite{delorme2004eeglab}. The EEG data was further filtered using a zero-phase windowed FIR filter with a Hamming window in the delta band (0.5-3 Hz). A total of 21 electrodes from the motor cortex region ('F3', 'Fz', 'F4', 'FC5', 'FC1', 'FC2', 'FC6', 'C3', 'Cz', 'C4', 'CP5', 'CP1', 'CP2', 'CP6', 'P7', 'P3', 'Pz','P4'), and occipital region ('O1', 'Oz', 'O2') of the brain were considered. Each EEG channel data is standardized using z-normalization technique.

The kinematic data was smoothed using a zero-phase low pass FIR filter with a cut-off frequency of 2 Hz. The filtered data was further normalized between 0 and 1 using the min-max normalization technique. This data was further down-sampled to 100 Hz in order to match the sampling rate of the EEG data.

\begin{figure}[b]
	\centering
	\includegraphics[width=0.3\textwidth]{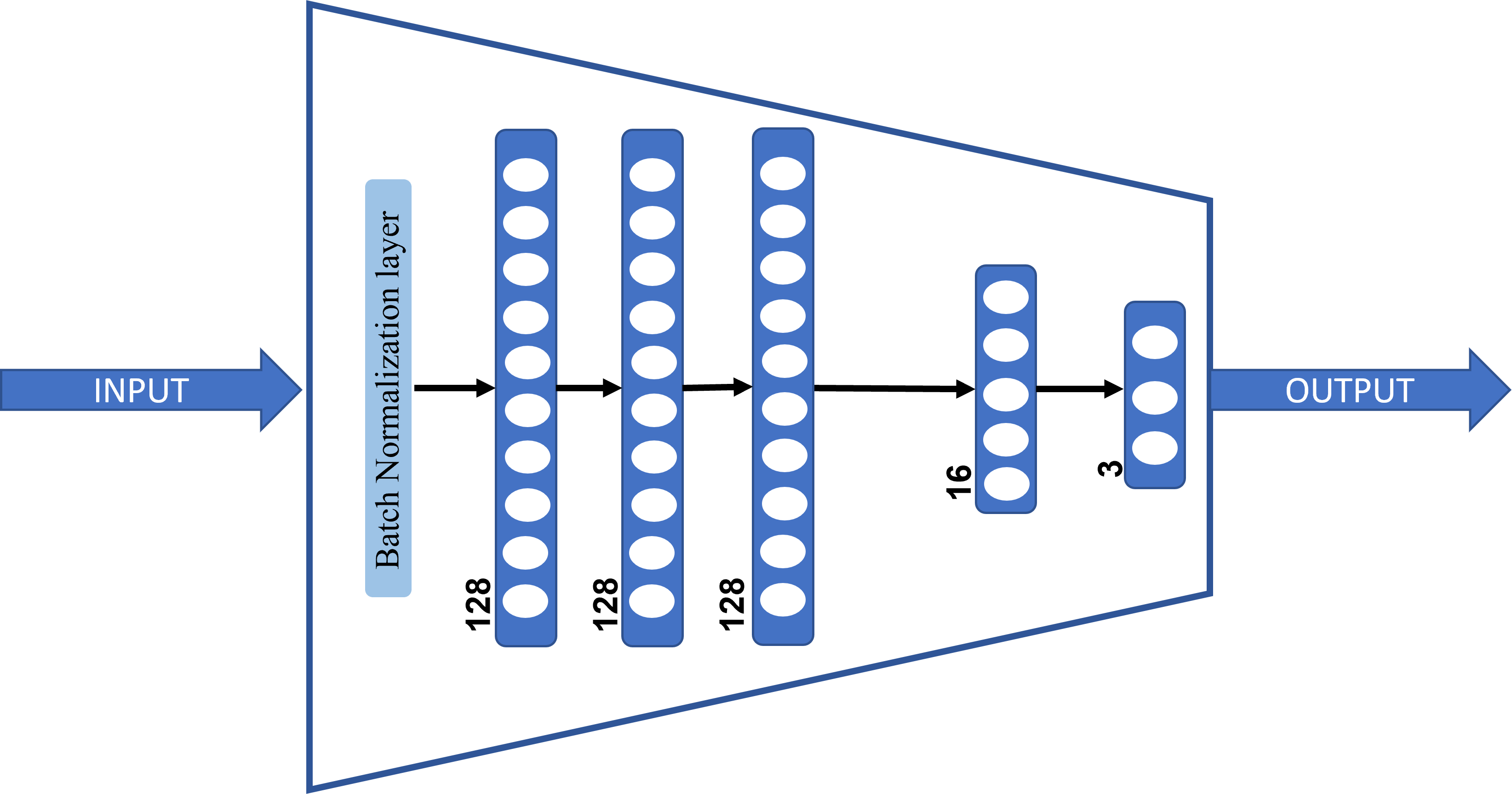}
	\caption{PreMovNet-I based on Multi-layer Perceptron model structure for 3D hand kinematics decoding.\vspace{-0.25 cm}}
	\label{01_MLP_model}
\end{figure}

\begin{figure}[b]
	\centering
	\includegraphics[width=0.37\textwidth]{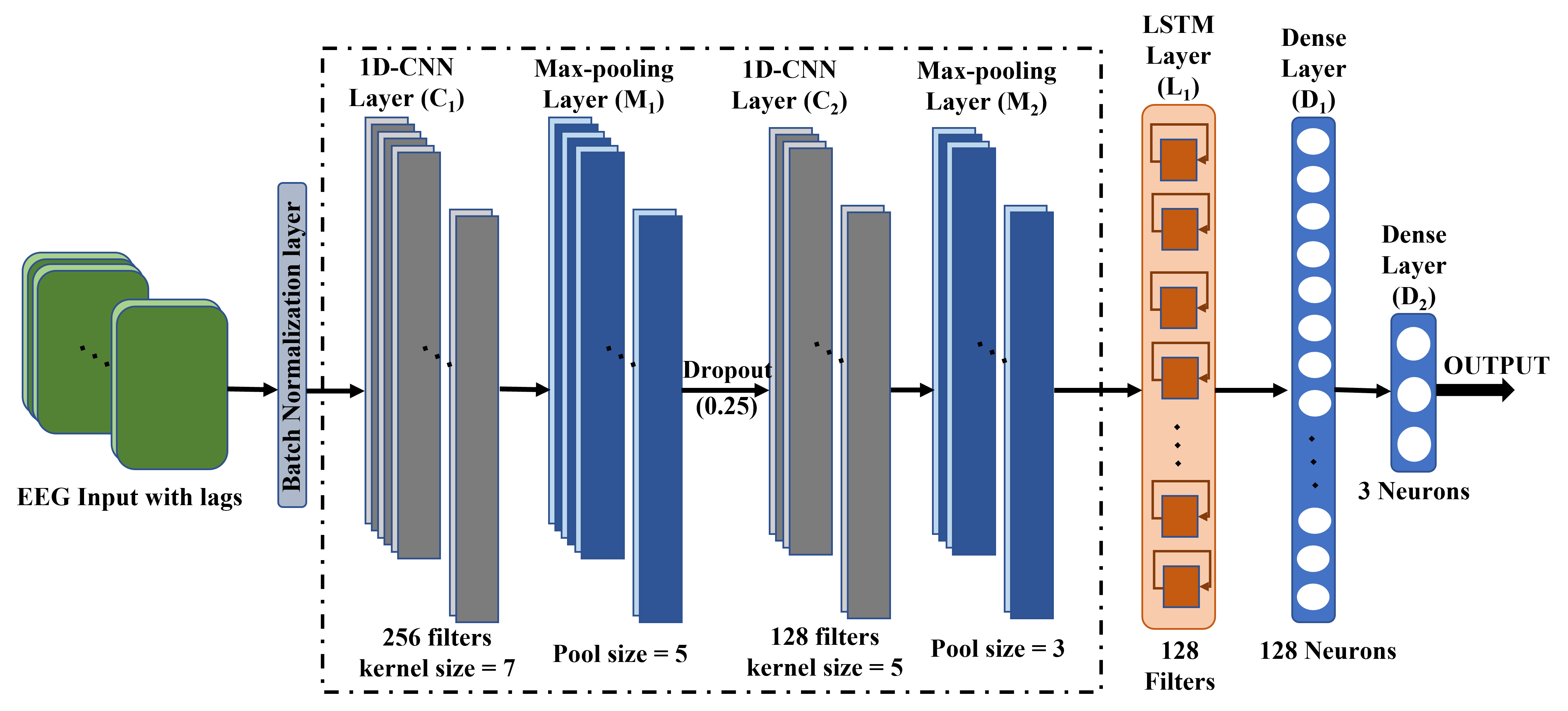}
	\caption{PreMovNet-II based on CNN-LSTM architecture for 3D hand kinematics decoding.}
	\label{02_PreMovNet}
\end{figure}

\subsection{Data preparation}
The kinematics data segment was initiated from the movement onset till the subject puts their hand in the resting position. It has been observed that the brain activity appears in the motor cortex region prior to the movement onset \cite{pancholi2022source}. We consolidated the motor cortex activity information prior to the motor activity by incorporating the time lags of the selected EEG channels from the movement onset. In particular, time lags of 15 samples, 20 samples, 25 samples, 30 samples, and 35 samples were included in this study. With sampling rate as $100$ Hz, EEG segment corresponding to time lags of 15 samples will be (-150ms to 0 ms) where 0ms is taken from the movement onset. For each kinematic vector, an EEG input matrix of dimension $T\times(L*N)$ was generated where $T$ is the data segment, $L$ is the time lag, and $N$ is total EEG channels selected.
\begin{table*}[!t]
\centering
\caption{PCC analysis of (a) mLR, (b) PreMovNet-I (MLP-based), and (c) PreMovNet-II (CNN-LSTM based) neural decoders with time lags of 150 ms, 200 ms, 250 ms, 300 ms and 350 ms.}
\scalebox{0.5}{
\centering
\begin{tabular}{|c|c|ccc|ccc|ccc|ccc|ccc|}
\hline
                                       &                                      & \multicolumn{3}{c|}{\textbf{150 ms}}                                                            & \multicolumn{3}{c|}{\textbf{200 ms}}                                                            & \multicolumn{3}{c|}{\textbf{250 ms}}                                                                                                          & \multicolumn{3}{c|}{\textbf{300 ms}}                                                            & \multicolumn{3}{c|}{\textbf{350 ms}}                                                            \\ \cline{3-17} 
\multirow{-2}{*}{\textbf{PARTICIPANT}} & \multirow{-2}{*}{\textbf{Direction}} & \multicolumn{1}{c|}{\textbf{mLR}} & \multicolumn{1}{c|}{\textbf{PreMovNet-I}}    & \textbf{PreMovNet-II} & \multicolumn{1}{c|}{\textbf{mLR}} & \multicolumn{1}{c|}{\textbf{PreMovNet-I}}    & \textbf{PreMovNet-II} & \multicolumn{1}{c|}{\textbf{mLR}} & \multicolumn{1}{c|}{\textbf{PreMovNet-I}}                            & \textbf{PreMovNet-II}                       & \multicolumn{1}{c|}{\textbf{mLR}} & \multicolumn{1}{c|}{\textbf{PreMovNet-I}}    & \textbf{PreMovNet-II} & \multicolumn{1}{c|}{\textbf{mLR}} & \multicolumn{1}{c|}{\textbf{PreMovNet-I}}    & \textbf{PreMovNet-II} \\ \hline
                                       & \textbf{x}                           & \multicolumn{1}{c|}{0.6692}       & \multicolumn{1}{c|}{0.7249}          & \textbf{0.7350}   & \multicolumn{1}{c|}{0.6537}       & \multicolumn{1}{c|}{0.7335}          & \textbf{0.7397}   & \multicolumn{1}{c|}{0.6542}       & \multicolumn{1}{c|}{0.8035}                                  & \textbf{0.8131}                         & \multicolumn{1}{c|}{0.6410}       & \multicolumn{1}{c|}{0.7729}          & \textbf{0.7926}   & \multicolumn{1}{c|}{0.6396}       & \multicolumn{1}{c|}{\textbf{0.7955}} & 0.7777            \\ \cline{2-17} 
                                       & \textbf{y}                           & \multicolumn{1}{c|}{0.6621}       & \multicolumn{1}{c|}{0.7122}          & \textbf{0.7284}   & \multicolumn{1}{c|}{0.6470}       & \multicolumn{1}{c|}{0.7198}          & \textbf{0.7336}   & \multicolumn{1}{c|}{0.6462}       & \multicolumn{1}{c|}{0.7930}                                  & \textbf{0.8008}                         & \multicolumn{1}{c|}{0.6350}       & \multicolumn{1}{c|}{0.7639}          & \textbf{0.7893}   & \multicolumn{1}{c|}{0.6335}       & \multicolumn{1}{c|}{\textbf{0.7767}} & 0.7596            \\ \cline{2-17} 
\multirow{-3}{*}{\textbf{P1}}          & \textbf{z}                           & \multicolumn{1}{c|}{0.4453}       & \multicolumn{1}{c|}{\textbf{0.6134}} & 0.5948            & \multicolumn{1}{c|}{0.4438}       & \multicolumn{1}{c|}{0.6506}          & \textbf{0.6671}   & \multicolumn{1}{c|}{0.4349}       & \multicolumn{1}{c|}{0.7116}                                  & \textbf{0.7164}                         & \multicolumn{1}{c|}{0.4360}       & \multicolumn{1}{c|}{0.6921}          & \textbf{0.7186}   & \multicolumn{1}{c|}{0.3984}       & \multicolumn{1}{c|}{0.6949}          & \textbf{0.7045}   \\ \hline
                                       & \textbf{x}                           & \multicolumn{1}{c|}{0.4527}       & \multicolumn{1}{c|}{0.6915}          & \textbf{0.6956}   & \multicolumn{1}{c|}{0.4612}       & \multicolumn{1}{c|}{0.6794}          & \textbf{0.6899}   & \multicolumn{1}{c|}{0.4369}       & \multicolumn{1}{c|}{0.6963}                                  & \textbf{0.7608}                         & \multicolumn{1}{c|}{0.4700}       & \multicolumn{1}{c|}{0.7181}          & \textbf{0.7763}   & \multicolumn{1}{c|}{0.4460}       & \multicolumn{1}{c|}{0.7398}          & \textbf{0.7465}   \\ \cline{2-17} 
                                       & \textbf{y}                           & \multicolumn{1}{c|}{0.4513}       & \multicolumn{1}{c|}{\textbf{0.7114}} & 0.7037            & \multicolumn{1}{c|}{0.4641}       & \multicolumn{1}{c|}{0.6943}          & \textbf{0.7049}   & \multicolumn{1}{c|}{0.4466}       & \multicolumn{1}{c|}{0.7056}                                  & \textbf{0.7581}                         & \multicolumn{1}{c|}{0.4744}       & \multicolumn{1}{c|}{0.7329}          & \textbf{0.7758}   & \multicolumn{1}{c|}{0.4561}       & \multicolumn{1}{c|}{\textbf{0.7500}} & 0.7438            \\ \cline{2-17} 
\multirow{-3}{*}{\textbf{P2}}          & \textbf{z}                           & \multicolumn{1}{c|}{0.2333}       & \multicolumn{1}{c|}{\textbf{0.4286}} & 0.3945            & \multicolumn{1}{c|}{0.2695}       & \multicolumn{1}{c|}{0.5251}          & \textbf{0.5301}   & \multicolumn{1}{c|}{0.2810}       & \multicolumn{1}{c|}{0.5072}                                  & \textbf{0.5112}                         & \multicolumn{1}{c|}{0.3077}       & \multicolumn{1}{c|}{\textbf{0.5599}} & 0.5122            & \multicolumn{1}{c|}{0.3056}       & \multicolumn{1}{c|}{\textbf{0.5492}} & 0.5198            \\ \hline
                                       & \textbf{x}                           & \multicolumn{1}{c|}{0.5157}       & \multicolumn{1}{c|}{0.6392}          & \textbf{0.6605}   & \multicolumn{1}{c|}{0.5877}       & \multicolumn{1}{c|}{0.6768}          & \textbf{0.6789}   & \multicolumn{1}{c|}{0.6028}       & \multicolumn{1}{c|}{0.6972}                                  & \textbf{0.7601}                         & \multicolumn{1}{c|}{0.5510}       & \multicolumn{1}{c|}{\textbf{0.6950}} & 0.6903            & \multicolumn{1}{c|}{0.5985}       & \multicolumn{1}{c|}{0.7329}          & \textbf{0.7478}   \\ \cline{2-17} 
                                       & \textbf{y}                           & \multicolumn{1}{c|}{0.5273}       & \multicolumn{1}{c|}{0.6381}          & \textbf{0.6703}   & \multicolumn{1}{c|}{0.6067}       & \multicolumn{1}{c|}{\textbf{0.6812}} & 0.6794            & \multicolumn{1}{c|}{0.6165}       & \multicolumn{1}{c|}{0.6965}                                  & \textbf{0.7604}                         & \multicolumn{1}{c|}{0.5661}       & \multicolumn{1}{c|}{\textbf{0.7011}} & 0.6732            & \multicolumn{1}{c|}{0.6100}       & \multicolumn{1}{c|}{0.7391}          & \textbf{0.7469}   \\ \cline{2-17} 
\multirow{-3}{*}{\textbf{P3}}          & \textbf{z}                           & \multicolumn{1}{c|}{0.4845}       & \multicolumn{1}{c|}{\textbf{0.6361}} & 0.6221            & \multicolumn{1}{c|}{0.5158}       & \multicolumn{1}{c|}{0.6433}          & \textbf{0.6457}   & \multicolumn{1}{c|}{0.4917}       & \multicolumn{1}{c|}{0.6402}                                  & \textbf{0.6855}                         & \multicolumn{1}{c|}{0.4304}       & \multicolumn{1}{c|}{0.6334}          & \textbf{0.6620}   & \multicolumn{1}{c|}{0.4437}       & \multicolumn{1}{c|}{0.6577}          & \textbf{0.6885}   \\ \hline
                                       & \textbf{x}                           & \multicolumn{1}{c|}{0.6720}       & \multicolumn{1}{c|}{0.8263}          & \textbf{0.8352}   & \multicolumn{1}{c|}{0.6689}       & \multicolumn{1}{c|}{0.8343}          & \textbf{0.8628}   & \multicolumn{1}{c|}{0.6681}       & \multicolumn{1}{c|}{\textbf{0.8439}}                         & 0.8249                                  & \multicolumn{1}{c|}{0.6512}       & \multicolumn{1}{c|}{\textbf{0.8183}} & 0.8040            & \multicolumn{1}{c|}{0.6389}       & \multicolumn{1}{c|}{\textbf{0.7683}} & 0.7559            \\ \cline{2-17} 
                                       & \textbf{y}                           & \multicolumn{1}{c|}{0.6972}       & \multicolumn{1}{c|}{0.8438}          & \textbf{0.8543}   & \multicolumn{1}{c|}{0.6941}       & \multicolumn{1}{c|}{0.8549}          & \textbf{0.8829}   & \multicolumn{1}{c|}{0.6913}       & \multicolumn{1}{c|}{\textbf{0.8587}}                         & 0.8473                                  & \multicolumn{1}{c|}{0.6736}       & \multicolumn{1}{c|}{\textbf{0.8353}} & 0.8220            & \multicolumn{1}{c|}{0.6604}       & \multicolumn{1}{c|}{\textbf{0.7913}} & 0.7688            \\ \cline{2-17} 
\multirow{-3}{*}{\textbf{P4}}          & \textbf{z}                           & \multicolumn{1}{c|}{0.4784}       & \multicolumn{1}{c|}{0.6704}          & \textbf{0.7228}   & \multicolumn{1}{c|}{0.4715}       & \multicolumn{1}{c|}{\textbf{0.7280}} & 0.6966            & \multicolumn{1}{c|}{0.4573}       & \multicolumn{1}{c|}{\textbf{0.7023}}                         & 0.7020                                  & \multicolumn{1}{c|}{0.4317}       & \multicolumn{1}{c|}{\textbf{0.7167}} & 0.7048            & \multicolumn{1}{c|}{0.4086}       & \multicolumn{1}{c|}{0.6814}          & \textbf{0.7205}   \\ \hline
                                       & \textbf{x}                           & \multicolumn{1}{c|}{0.5274}       & \multicolumn{1}{c|}{0.7311}          & \textbf{0.7992}   & \multicolumn{1}{c|}{0.5380}       & \multicolumn{1}{c|}{0.7858}          & \textbf{0.8108}   & \multicolumn{1}{c|}{0.5408}       & \multicolumn{1}{c|}{0.7810}                                  & \textbf{0.7924}                         & \multicolumn{1}{c|}{0.5304}       & \multicolumn{1}{c|}{0.7602}          & \textbf{0.7870}   & \multicolumn{1}{c|}{0.5016}       & \multicolumn{1}{c|}{0.7606}          & \textbf{0.8005}   \\ \cline{2-17} 
                                       & \textbf{y}                           & \multicolumn{1}{c|}{0.5422}       & \multicolumn{1}{c|}{0.7667}          & \textbf{0.8171}   & \multicolumn{1}{c|}{0.5588}       & \multicolumn{1}{c|}{0.8150}          & \textbf{0.8306}   & \multicolumn{1}{c|}{0.5633}       & \multicolumn{1}{c|}{0.8089}                                  & \textbf{0.8185}                         & \multicolumn{1}{c|}{0.5548}       & \multicolumn{1}{c|}{\textbf{0.8120}} & 0.8106            & \multicolumn{1}{c|}{0.5294}       & \multicolumn{1}{c|}{0.7990}          & \textbf{0.8383}   \\ \cline{2-17} 
\multirow{-3}{*}{\textbf{P5}}          & \textbf{z}                           & \multicolumn{1}{c|}{0.5154}       & \multicolumn{1}{c|}{0.6162}          & \textbf{0.6275}   & \multicolumn{1}{c|}{0.5279}       & \multicolumn{1}{c|}{0.6229}          & \textbf{0.6469}   & \multicolumn{1}{c|}{0.5052}       & \multicolumn{1}{c|}{\textbf{0.6578}}                         & 0.6482                                  & \multicolumn{1}{c|}{0.4786}       & \multicolumn{1}{c|}{\textbf{0.6658}} & 0.6381            & \multicolumn{1}{c|}{0.4499}       & \multicolumn{1}{c|}{0.6409}          & \textbf{0.6554}   \\ \hline
                                       & \textbf{x}                           & \multicolumn{1}{c|}{0.3589}       & \multicolumn{1}{c|}{0.7838}          & \textbf{0.8033}   & \multicolumn{1}{c|}{0.3744}       & \multicolumn{1}{c|}{0.7708}          & \textbf{0.8090}   & \multicolumn{1}{c|}{0.3641}       & \multicolumn{1}{c|}{0.8046}                                  & \textbf{0.8380}                         & \multicolumn{1}{c|}{0.4541}       & \multicolumn{1}{c|}{0.7740}          & \textbf{0.8288}   & \multicolumn{1}{c|}{0.4173}       & \multicolumn{1}{c|}{\textbf{0.7922}} & 0.7655            \\ \cline{2-17} 
                                       & \textbf{y}                           & \multicolumn{1}{c|}{0.3711}       & \multicolumn{1}{c|}{0.7602}          & \textbf{0.7860}   & \multicolumn{1}{c|}{0.3873}       & \multicolumn{1}{c|}{0.7529}          & \textbf{0.7951}   & \multicolumn{1}{c|}{0.3783}       & \multicolumn{1}{c|}{0.7888}                                  & \textbf{0.8308}                         & \multicolumn{1}{c|}{0.4606}       & \multicolumn{1}{c|}{0.7713}          & \textbf{0.8118}   & \multicolumn{1}{c|}{0.4263}       & \multicolumn{1}{c|}{\textbf{0.7735}} & 0.7597            \\ \cline{2-17} 
\multirow{-3}{*}{\textbf{P6}}          & \textbf{z}                           & \multicolumn{1}{c|}{0.2968}       & \multicolumn{1}{c|}{\textbf{0.5971}} & 0.5612            & \multicolumn{1}{c|}{0.3133}       & \multicolumn{1}{c|}{0.5822}          & \textbf{0.5939}   & \multicolumn{1}{c|}{0.3166}       & \multicolumn{1}{c|}{\textbf{0.6001}}                         & 0.5752                                  & \multicolumn{1}{c|}{0.3255}       & \multicolumn{1}{c|}{\textbf{0.6137}} & 0.6023            & \multicolumn{1}{c|}{0.3184}       & \multicolumn{1}{c|}{0.6217}          & \textbf{0.6383}   \\ \hline
                                       & \textbf{x}                           & \multicolumn{1}{c|}{0.4059}       & \multicolumn{1}{c|}{0.6981}          & \textbf{0.7380}   & \multicolumn{1}{c|}{0.4384}       & \multicolumn{1}{c|}{0.7168}          & \textbf{0.7826}   & \multicolumn{1}{c|}{0.4030}       & \multicolumn{1}{c|}{0.7158}                                  & \textbf{0.7791}                         & \multicolumn{1}{c|}{0.3652}       & \multicolumn{1}{c|}{0.7587}          & \textbf{0.7846}   & \multicolumn{1}{c|}{0.3858}       & \multicolumn{1}{c|}{0.7889}          & \textbf{0.7902}   \\ \cline{2-17} 
                                       & \textbf{y}                           & \multicolumn{1}{c|}{0.4078}       & \multicolumn{1}{c|}{0.7082}          & \textbf{0.7557}   & \multicolumn{1}{c|}{0.4373}       & \multicolumn{1}{c|}{0.7409}          & \textbf{0.8057}   & \multicolumn{1}{c|}{0.3951}       & \multicolumn{1}{c|}{0.7330}                                  & \textbf{0.8015}                         & \multicolumn{1}{c|}{0.3497}       & \multicolumn{1}{c|}{0.7714}          & \textbf{0.8045}   & \multicolumn{1}{c|}{0.3676}       & \multicolumn{1}{c|}{0.7964}          & \textbf{0.8083}   \\ \cline{2-17} 
\multirow{-3}{*}{\textbf{P7}}          & \textbf{z}                           & \multicolumn{1}{c|}{0.4347}       & \multicolumn{1}{c|}{\textbf{0.5484}} & 0.4968            & \multicolumn{1}{c|}{0.4471}       & \multicolumn{1}{c|}{0.5573}          & \textbf{0.5617}   & \multicolumn{1}{c|}{0.4191}       & \multicolumn{1}{c|}{\textbf{0.5602}}                         & 0.5386                                  & \multicolumn{1}{c|}{0.4041}       & \multicolumn{1}{c|}{\textbf{0.5659}} & 0.5580            & \multicolumn{1}{c|}{0.3809}       & \multicolumn{1}{c|}{0.5847}          & \textbf{0.6007}   \\ \hline
                                       & \textbf{x}                           & \multicolumn{1}{c|}{0.4956}       & \multicolumn{1}{c|}{0.7650}          & \textbf{0.7925}   & \multicolumn{1}{c|}{0.5218}       & \multicolumn{1}{c|}{0.6919}          & \textbf{0.7262}   & \multicolumn{1}{c|}{0.6095}       & \multicolumn{1}{c|}{0.8012}                                  & \textbf{0.8273}                         & \multicolumn{1}{c|}{0.6601}       & \multicolumn{1}{c|}{\textbf{0.8174}} & 0.8078            & \multicolumn{1}{c|}{0.6505}       & \multicolumn{1}{c|}{0.8032}          & \textbf{0.8343}   \\ \cline{2-17} 
                                       & \textbf{y}                           & \multicolumn{1}{c|}{0.5028}       & \multicolumn{1}{c|}{0.7615}          & \textbf{0.7989}   & \multicolumn{1}{c|}{0.5091}       & \multicolumn{1}{c|}{0.7007}          & \textbf{0.7299}   & \multicolumn{1}{c|}{0.6078}       & \multicolumn{1}{c|}{0.8081}                                  & \textbf{0.8228}                         & \multicolumn{1}{c|}{0.6534}       & \multicolumn{1}{c|}{\textbf{0.8174}} & 0.8026            & \multicolumn{1}{c|}{0.6419}       & \multicolumn{1}{c|}{0.8017}          & \textbf{0.8356}   \\ \cline{2-17} 
\multirow{-3}{*}{\textbf{P8}}          & \textbf{z}                           & \multicolumn{1}{c|}{0.2775}       & \multicolumn{1}{c|}{\textbf{0.6320}} & 0.6275            & \multicolumn{1}{c|}{0.1498}       & \multicolumn{1}{c|}{0.5701}          & \textbf{0.5744}   & \multicolumn{1}{c|}{0.2740}       & \multicolumn{1}{c|}{\textbf{0.7039}}                         & 0.6741                                  & \multicolumn{1}{c|}{0.2867}       & \multicolumn{1}{c|}{0.6925}          & \textbf{0.7115}   & \multicolumn{1}{c|}{0.2827}       & \multicolumn{1}{c|}{\textbf{0.6908}} & 0.6823            \\ \hline
                                       & \textbf{x}                           & \multicolumn{1}{c|}{0.4903}       & \multicolumn{1}{c|}{0.7545}          & \textbf{0.7868}   & \multicolumn{1}{c|}{0.5221}       & \multicolumn{1}{c|}{\textbf{0.7958}} & 0.7887            & \multicolumn{1}{c|}{0.5440}       & \multicolumn{1}{c|}{0.7699}                                  & \textbf{0.8223}                         & \multicolumn{1}{c|}{0.5292}       & \multicolumn{1}{c|}{\textbf{0.8024}} & 0.7903            & \multicolumn{1}{c|}{0.5441}       & \multicolumn{1}{c|}{\textbf{0.8108}} & 0.7938            \\ \cline{2-17} 
                                       & \textbf{y}                           & \multicolumn{1}{c|}{0.4877}       & \multicolumn{1}{c|}{0.7605}          & \textbf{0.7946}   & \multicolumn{1}{c|}{0.5207}       & \multicolumn{1}{c|}{0.7970}          & \textbf{0.7978}   & \multicolumn{1}{c|}{0.5429}       & \multicolumn{1}{c|}{0.7758}                                  & \textbf{0.8185}                         & \multicolumn{1}{c|}{0.5305}       & \multicolumn{1}{c|}{\textbf{0.8039}} & 0.7926            & \multicolumn{1}{c|}{0.5469}       & \multicolumn{1}{c|}{\textbf{0.8137}} & 0.7959            \\ \cline{2-17} 
\multirow{-3}{*}{\textbf{P9}}          & \textbf{z}                           & \multicolumn{1}{c|}{0.3442}       & \multicolumn{1}{c|}{\textbf{0.5782}} & 0.5141            & \multicolumn{1}{c|}{0.3436}       & \multicolumn{1}{c|}{\textbf{0.5994}} & 0.5408            & \multicolumn{1}{c|}{0.3631}       & \multicolumn{1}{c|}{0.5667}                                  & \textbf{0.5735}                         & \multicolumn{1}{c|}{0.3532}       & \multicolumn{1}{c|}{\textbf{0.6155}} & 0.5194            & \multicolumn{1}{c|}{0.3498}       & \multicolumn{1}{c|}{\textbf{0.6458}} & 0.5861            \\ \hline
                                       & \textbf{x}                           & \multicolumn{1}{c|}{0.6767}       & \multicolumn{1}{c|}{0.8200}          & \textbf{0.8393}   & \multicolumn{1}{c|}{0.3779}       & \multicolumn{1}{c|}{0.7508}          & \textbf{0.7984}   & \multicolumn{1}{c|}{0.3714}       & \multicolumn{1}{c|}{0.7629}                                  & \textbf{0.8164}                         & \multicolumn{1}{c|}{0.3860}       & \multicolumn{1}{c|}{0.7417}          & \textbf{0.7821}   & \multicolumn{1}{c|}{0.4483}       & \multicolumn{1}{c|}{0.7609}          & \textbf{0.7978}   \\ \cline{2-17} 
                                       & \textbf{y}                           & \multicolumn{1}{c|}{0.7040}       & \multicolumn{1}{c|}{0.8337}          & \textbf{0.8565}   & \multicolumn{1}{c|}{0.3989}       & \multicolumn{1}{c|}{0.7629}          & \textbf{0.8149}   & \multicolumn{1}{c|}{0.3943}       & \multicolumn{1}{c|}{0.7755}                                  & \textbf{0.8347}                         & \multicolumn{1}{c|}{0.4091}       & \multicolumn{1}{c|}{0.7525}          & \textbf{0.7949}   & \multicolumn{1}{c|}{0.4817}       & \multicolumn{1}{c|}{0.7773}          & \textbf{0.8085}   \\ \cline{2-17} 
\multirow{-3}{*}{\textbf{P10}}         & \textbf{z}                           & \multicolumn{1}{c|}{0.3393}       & \multicolumn{1}{c|}{\textbf{0.6634}} & 0.6376            & \multicolumn{1}{c|}{0.2487}       & \multicolumn{1}{c|}{\textbf{0.6374}} & 0.6359            & \multicolumn{1}{c|}{0.2465}       & \multicolumn{1}{c|}{0.6426}                                  & \textbf{0.6629}                         & \multicolumn{1}{c|}{0.1981}       & \multicolumn{1}{c|}{0.6228}          & \textbf{0.6384}   & \multicolumn{1}{c|}{0.2053}       & \multicolumn{1}{c|}{\textbf{0.6503}} & 0.6182            \\ \hline
                                       & \textbf{x}                           & \multicolumn{1}{c|}{0.5094}       & \multicolumn{1}{c|}{0.7233}          & \textbf{0.8140}   & \multicolumn{1}{c|}{0.4823}       & \multicolumn{1}{c|}{0.7367}          & \textbf{0.7677}   & \multicolumn{1}{c|}{0.4992}       & \multicolumn{1}{c|}{0.7669}                                  & \textbf{0.7800}                         & \multicolumn{1}{c|}{0.4530}       & \multicolumn{1}{c|}{0.7317}          & \textbf{0.7567}   & \multicolumn{1}{c|}{0.4423}       & \multicolumn{1}{c|}{0.7358}          & \textbf{0.7566}   \\ \cline{2-17} 
                                       & \textbf{y}                           & \multicolumn{1}{c|}{0.5342}       & \multicolumn{1}{c|}{0.7531}          & \textbf{0.8360}   & \multicolumn{1}{c|}{0.5131}       & \multicolumn{1}{c|}{0.7747}          & \textbf{0.7950}   & \multicolumn{1}{c|}{0.5203}       & \multicolumn{1}{c|}{0.7873}                                  & \textbf{0.8048}                         & \multicolumn{1}{c|}{0.4801}       & \multicolumn{1}{c|}{0.7601}          & \textbf{0.7853}   & \multicolumn{1}{c|}{0.4663}       & \multicolumn{1}{c|}{0.7608}          & \textbf{0.7841}   \\ \cline{2-17} 
\multirow{-3}{*}{\textbf{P11}}         & \textbf{z}                           & \multicolumn{1}{c|}{0.3683}       & \multicolumn{1}{c|}{\textbf{0.4867}} & 0.4742            & \multicolumn{1}{c|}{0.3944}       & \multicolumn{1}{c|}{\textbf{0.5354}} & 0.4774            & \multicolumn{1}{c|}{0.4250}       & \multicolumn{1}{c|}{\textbf{0.5378}}                         & 0.4801                                  & \multicolumn{1}{c|}{0.4265}       & \multicolumn{1}{c|}{\textbf{0.5612}} & 0.5091            & \multicolumn{1}{c|}{0.4547}       & \multicolumn{1}{c|}{0.5678}          & \textbf{0.5853}   \\ \hline
                                       & \textbf{x}                           & \multicolumn{1}{c|}{0.3132}       & \multicolumn{1}{c|}{0.6455}          & \textbf{0.6736}   & \multicolumn{1}{c|}{0.3429}       & \multicolumn{1}{c|}{0.5996}          & \textbf{0.6873}   & \multicolumn{1}{c|}{0.3182}       & \multicolumn{1}{c|}{0.5976}                                  & \textbf{0.6750}                         & \multicolumn{1}{c|}{0.3159}       & \multicolumn{1}{c|}{0.6096}          & \textbf{0.6527}   & \multicolumn{1}{c|}{0.3482}       & \multicolumn{1}{c|}{0.6428}          & \textbf{0.6811}   \\ \cline{2-17} 
                                       & \textbf{y}                           & \multicolumn{1}{c|}{0.3328}       & \multicolumn{1}{c|}{0.6616}          & \textbf{0.6778}   & \multicolumn{1}{c|}{0.3688}       & \multicolumn{1}{c|}{0.5966}          & \textbf{0.6991}   & \multicolumn{1}{c|}{0.3442}       & \multicolumn{1}{c|}{0.5953}                                  & \textbf{0.6904}                         & \multicolumn{1}{c|}{0.3482}       & \multicolumn{1}{c|}{0.6154}          & \textbf{0.6522}   & \multicolumn{1}{c|}{0.3742}       & \multicolumn{1}{c|}{0.6504}          & \textbf{0.6911}   \\ \cline{2-17} 
\multirow{-3}{*}{\textbf{P12}}         & \textbf{z}                           & \multicolumn{1}{c|}{0.3486}       & \multicolumn{1}{c|}{\textbf{0.4625}} & 0.4169            & \multicolumn{1}{c|}{0.3688}       & \multicolumn{1}{c|}{\textbf{0.4368}} & 0.4161            & \multicolumn{1}{c|}{0.3868}       & \multicolumn{1}{c|}{0.4375}                                  & \textbf{0.4377}                         & \multicolumn{1}{c|}{0.3904}       & \multicolumn{1}{c|}{\textbf{0.4560}} & 0.4052            & \multicolumn{1}{c|}{0.4034}       & \multicolumn{1}{c|}{\textbf{0.4619}} & 0.4318            \\ \hline \hline
                                       & \textbf{x}                           & \multicolumn{1}{c|}{0.5072}       & \multicolumn{1}{c|}{0.7336}          & \textbf{0.7644}   & \multicolumn{1}{c|}{0.4974}       & \multicolumn{1}{c|}{0.7310}          & \textbf{0.7618}   & \multicolumn{1}{c|}{0.5010}       & \multicolumn{1}{c|}{0.7534}                                  & \cellcolor[HTML]{9AFF99}\textbf{0.7908} & \multicolumn{1}{c|}{0.5006}       & \multicolumn{1}{c|}{0.7500}          & \textbf{0.7711}   & \multicolumn{1}{c|}{0.5051}       & \multicolumn{1}{c|}{0.7610}          & \textbf{0.7706}   \\ \cline{2-17} 
                                       & \textbf{y}                           & \multicolumn{1}{c|}{0.5184}       & \multicolumn{1}{c|}{0.7426}          & \textbf{0.7733}   & \multicolumn{1}{c|}{0.5088}       & \multicolumn{1}{c|}{0.7409}          & \textbf{0.7724}   & \multicolumn{1}{c|}{0.5122}       & \multicolumn{1}{c|}{0.7605}                                  & \cellcolor[HTML]{9AFF99}\textbf{0.7990} & \multicolumn{1}{c|}{0.5113}       & \multicolumn{1}{c|}{0.7614}          & \textbf{0.7762}   & \multicolumn{1}{c|}{0.5162}       & \multicolumn{1}{c|}{0.7692}          & \textbf{0.7784}   \\ \cline{2-17}
\multirow{-3}{*}{\textbf{Average}}     & \textbf{z}                           & \multicolumn{1}{c|}{0.3805}       & \multicolumn{1}{c|}{\textbf{0.5778}} & 0.5575            & \multicolumn{1}{c|}{0.3745}       & \multicolumn{1}{c|}{\textbf{0.5907}} & 0.5822            & \multicolumn{1}{c|}{0.3834}       & \multicolumn{1}{c|}{\textbf{0.6056}} & 0.6005                                  & \multicolumn{1}{c|}{0.3724}       & \multicolumn{1}{c|}{\textbf{0.6163}} & 0.5983            & \multicolumn{1}{c|}{0.3668}       &  \cellcolor[HTML]{9AFF99}{\textbf{0.6206}} & 0.6193            \\ \hline
\end{tabular}
}
\label{tab:tab01}
\end{table*}
\vspace{-0.35cm}
\subsection{Multi-variable Linear Regression (mLR) model}
Multi-variable linear regression (mLR) based kinematic decoding is widely being used for brain-computer interface \cite{bradberry2010reconstructing,robinson2015adaptive,Sosnik2020}. The mLR model utilises multiple inputs to predict hand position directions. The input-output mapping for the mLR model is given as:
\begin{align}
H_x[t] =& \alpha_x+\sum_{n=1}^{N}\sum_{l=0}^{L}\beta^{(nl)}_{x}V_n[t-l]\\
H_y[t] =& \alpha_y+\sum_{n=1}^{N}\sum_{l=0}^{L}\beta^{(nl)}_{y}V_n[t-l]\\
H_z[t] =& \alpha_z+\sum_{n=1}^{N}\sum_{l=0}^{L}\beta^{(nl)}_{z}V_n[t-l]\\ \nonumber
\end{align}
the outputs, $H_x[t]$, $H_y[t]$, and $H_z[t]$ are the horizontal, vertical, and depth positions of the hand at time $t$, respectively. $V_n[t-l]$ is the standardized EEG potential at time lag $l$, where the number of time lags is varied from $0$ to $L$. The regression coefficients, $\alpha$ and $\beta$, are the coefficients of the mLR model.

\vspace{-0.25cm}

\subsection{Proposed PreMovNet-I: Multi-Layer Perceptron (MLP) based}
The multi-layer perceptron (MLP) based PreMovNet model is presented in Fig. \ref{01_MLP_model}. The model consists of six layers, including one batch normalisation layer, four dense layers, and one output layer. The first three dense layers consist of 128 neurons each, and the last dense layer consists of 16 neurons. The output layer consists of three neurons corresponding to the hand trajectory values in the x, y, and z directions as shown in Fig. \ref{01_MLP_model}.

\vspace{-0.25cm}

\subsection{Proposed PreMovNet-II: CNN-LSTM based}
In this Section, a CNN-LSTM based PreMovNet model, as shown in Fig. \ref{02_PreMovNet}, is presented for hand kinematics decoding for grasp and lift task. It consists of nine layers that include of batch normalisation layer, two convolution layers, two max-pooling layers, one dropout layer, one LSTM layer, and two dense layers. The batch-normalization layer is represented as $B_1$. The convolutional layer $C_1$ has a kernel size of $7$ and 256 filters, while $C_2$ layer has a kernel size of $5$ and 128 filters. The zero padding is done for both $C_1$ and $C_2$ layers so that the size of input and output remains the same. Each convolution layer is followed by a ReLu activation unit. $M_1$ and $M_2$ are max-pooling layers with a window size of $5$ and $3$, respectively. The layer, $D_1$ is a dropout layer with a 0.25 dropout rate. The layer $L_1$ is the LSTM layer with 128 cells, followed by the ReLU activation function. The last two layers, $D_1$ and $D_2$ are dense layers with 128 and 3 neurons, respectively. The last layer with three output neurons yields the predicted hand kinematics in the x, y, and z-directions as shown in Fig. \ref{02_PreMovNet}.

\subsection{Training and Evaluation}
With the aim of training and performance evaluation of the kinematics decoding models, the dataset was partitioned into separate training, validation, and test sets. The training set was used to train the models, while the validation set was used to tune the model hyper-parameters. The test set was used for the performance evaluation of the trained models. The adaptive moment estimation (Adam) optimization algorithm \cite{kingma2014adam} was employed with a mean squared error (MSE) loss function for training the deep learning neural decoders. The early stopping technique was adopted to avoid over-fitting of the neural decoders. Total 294 trials were utilised for each participant from the WAY-EEG-GAL dataset. Total trials were divided into three separate subsets for each subject: (a) training data consisting of 234 trials data samples; (b) validation data with 30 trials data samples; and (c) test data with 30 trials data samples. The performance of the trained models is evaluated using Pearson's correlation coefficient (PCC) between the predicted hand kinematics and measured kinematics data. It may be noted that the training step is computationally strenuous, while the estimation is rapid. Hence, the decoding model can be implemented to control external prostheses or exoskeletons/exosuits.

\begin{figure}[t]
	\centering
	\includegraphics[width=0.25\textwidth]{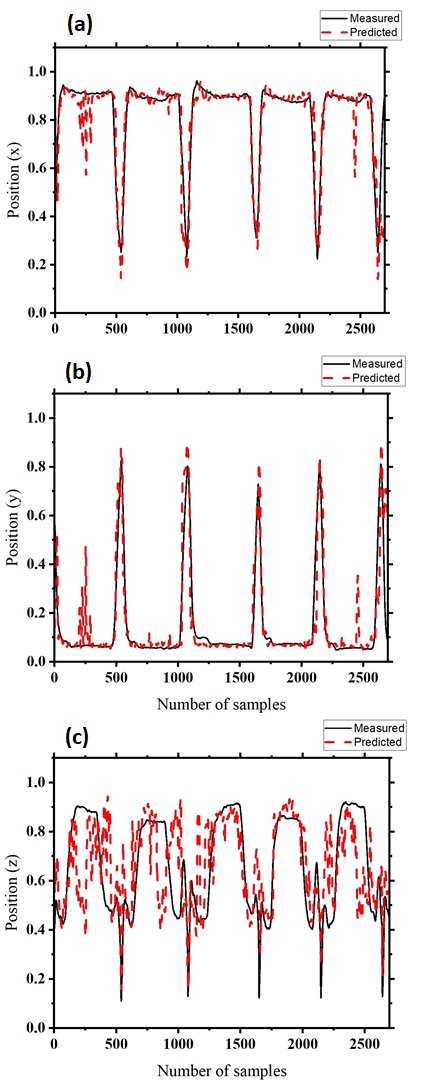}
	\caption{Trajectory estimation in x, y and z direction using the PreMovNet neural decoder is presented in (a)-(c) respectively.}
	\label{cnn_trajectory}
\end{figure}
\vspace{-0.15cm}
\section{Results}
PCC is utilised herein as a performance metric for measuring the efficiency of the neural decoder. In particular, PCC is evaluated at different time lags and compared the proposed neural decoders with the state-of-the-art mLR neural decoder.

The PCC analysis for the various neural decoders is presented in Table-\ref{tab:tab01} with varying time lags. The analysis is done for twelve subjects from the WAY-EEG-GAL dataset in all the x, y, and z directions. It may be observed that the proposed PreMovNet-I and PreMovNet-II perform reasonably better than the state-of-the-art mLR decoder. Additionally, the CNN-LSTM based PreMovNet-II performs better than MLP-based PreMovNet-I in the x and y-directions, while in the z-direction, the two proposed decoders have similar performance. In particular, CNN-LSTM based PreMovNet-II performs best with 250 ms time lag in x and y directions, while in the z-direction, the best performance is achieved by MLP-based PreMovNet-I at time lag of 350 ms. The average PCC of the CNN-LSTM based PreMovNet with 250 ms time lag is $0.7908\pm 0.043$, $0.7990\pm 0.042$ and $0.6005\pm 0.090$ in the x, y, and z-directions, respectively. The poor correlation in z-direction may be because of transient movement along it.

The predicted hand trajectory is additionally compared herein with measured trajectory. Fig. \ref{cnn_trajectory} (a)-(c) shows the measured and predicted hand trajectories of participant 04 for the CNN-LSTM based PreMovNet-II neural decoder in x, y, and z-directions, respectively with 200 ms lag. It may be noted that the estimated trajectories in x and y directions closely follow the measured ones. Poor correlation in z-direction results in less accurate estimated trajectory.
\vspace{-0.25cm}
\section{Conclusions and Future directions}
In this letter, two deep learning-based neural decoders are proposed for motion trajectory decoding (MTD) that make use of pre-movement EEG for efficient hand kinematics decoding. The inclusion of the neural motor information before the movement onset as an input feature improves the MTD ability of the proposed neural decoders. The proposed neural decoders, PreMovNet-I and PreMovNet-II, are compared with state-of-the-art mLR decoder on WAY-EEG-GAL dataset. The performance evaluation of the proposed decoders is presented using the Pearson correlation coefficient analysis. Additionally, the decoded hand trajectory is compared with the measured hand trajectory in x, y, and z-directions. A significant improvement in correlation and estimated trajectory is observed using deep-learning based PreMovNet when compared with traditional mLR based approach. The neural decoder for MTD can be selected based on performance and processing capabilities from the proposed models. The hand kinematics decoding may be employed for motor neurorehabilitation and power augmentation by controlling the external BCI devices such as exoskeletons/exosuits.

\section*{Acknowledgment}
The authors would like to thank Prof. Sitikantha Roy, Prof. Shubhendu Bhasin, and Prof. Sushma Santapuri from Indian Institute of Technology Delhi (IITD), and Dr. Suriya Prakash from All India Institute of Medical Sciences (AIIMS) Delhi for their discussion and constructive comments during the preparation of the manuscript.
\bibliographystyle{IEEEtran}
\bibliography{PreMovNet_v02}

\end{document}